\documentclass[preprint2]{aastex}

\shorttitle{The transiting exoplanet WASP-29b}
\shortauthors{Hellier et al.}

\begin{document}

\title{WASP-29b: A Saturn-sized transiting exoplanet}

\author{
Coel Hellier\altaffilmark{1}, %\email{ch@astro.keele.ac.uk}
D.R. Anderson\altaffilmark{1}, 
A. Collier Cameron\altaffilmark{2}, 
M. Gillon\altaffilmark{3}, 
M. Lendl\altaffilmark{4}, 
P.F.L. Maxted\altaffilmark{1}, 
D. Queloz\altaffilmark{4}, 
B. Smalley\altaffilmark{1}, 
A.H.M.J. Triaud\altaffilmark{4}, 
R.G. West\altaffilmark{5},
D.J.A. Brown\altaffilmark{2},
B. Enoch\altaffilmark{2}, 
T.A. Lister\altaffilmark{6}, 
F. Pepe\altaffilmark{4}, 
D. Pollacco\altaffilmark{7}, 
D. S\'egransan\altaffilmark{4}, 
S. Udry\altaffilmark{4}}

\altaffiltext{1}{Astrophysics Group, Keele University, Staffordshire, ST5 5BG, UK}
\altaffiltext{2}{School of Physics and Astronomy, University of St.\ Andrews, North Haugh,  Fife, KY16 9SS, UK}
\altaffiltext{3}{Institut d'Astrophysique et de G\'eophysique, Universit\'e de
Li\`ege, All\'ee du 6 Ao\^ut, 17, Bat. B5C, Li\`ege 1, Belgium,} 
\altaffiltext{4}{Observatoire astronomique de l'Universit\'e de Gen\`eve
51 ch. des Maillettes, 1290 Sauverny, Switzerland}
\altaffiltext{5}{Department of Physics and Astronomy, University of Leicester, Leicester, LE1 7RH, UK}
\altaffiltext{6}{Las Cumbres Observatory, 6740 Cortona Dr. Suite 102, Santa Barbara, CA 93117, USA}
\altaffiltext{7}{Astrophysics Research Centre, School of Mathematics \& Physics, Queen's University, University Road, Belfast, BT7 1NN, UK}

\begin{abstract}
We report the discovery of a Saturn-sized planet transiting a $V$ =
11.3, K4 dwarf star every 3.9 d. WASP-29b has a mass of 0.24\,$\pm$\,0.02 
M$_{\rm Jup}$ and a radius of 0.79\,$\pm$\,0.05 R$_{\rm Jup}$, making
it the smallest planet so far discovered by the WASP survey,
and the exoplanet most similar in mass and radius to Saturn.  
The host star WASP-29 has an above-Solar
metallicity and fits a possible correlation for Saturn-mass planets
such that planets with higher-metallicity host stars have higher core
masses and thus smaller radii.
\end{abstract}


\section{Introduction}
Searches for transiting exoplanets have now found more than
50 ``hot Jupiters'' with masses of $\sim$\,0.5--3 Jupiters.
At much smaller masses there are several transiting ``Neptunes'' 
(GJ\,436b, Gillon et\,al.\ 2007; HAT-P-11b, Bakos et\,al.\ 2010; \&\ 
Kepler-4b, Borucki et\,al.\ 2010) and ``super-Earths'' (GJ1214b, 
Charbonneau et\,al.\ 2009; CoRoT-7b, L\'eger et\,al.\ 2009).

By 2009 there were only two known transiting planets of Saturn-mass 
($\sim$\,0.3 M$_{\rm Jup}$), namely HD\,149026b (Sato et\,al.\ 2005) and 
HAT-P-12b (Hartman et\,al.\ 2009). In 2010
this number is growing fast, with near simultaneous announcements of 
WASP-29b (this paper), CoRoT-8b (Bord\'e et\,al.\ 2010),
WASP-21b (Bouchy et\,al.\ 2010) and HAT-P-18b and HAT-P-19b
(Hartman et\,al.\ 2010), giving rapidly increasing insight into 
planets of this mass range.  

\begin{figure}
\hspace*{-5mm}\includegraphics[width=9cm]{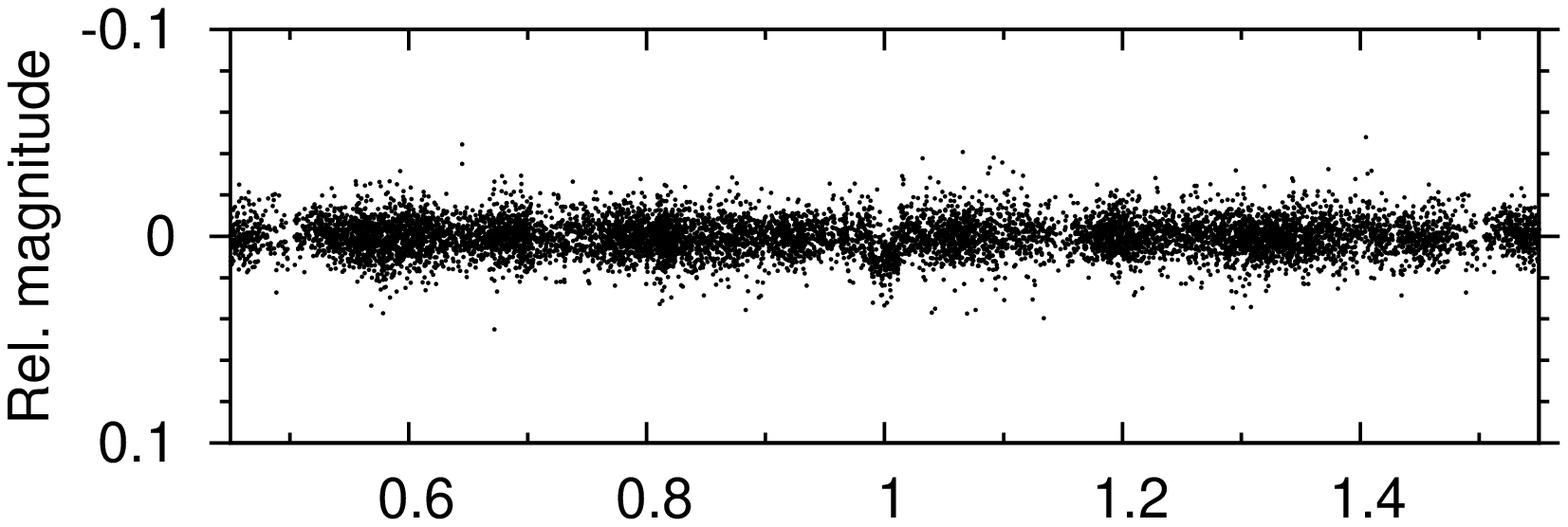}\\ [-2mm]
\hspace*{-5mm}\includegraphics[width=9cm]{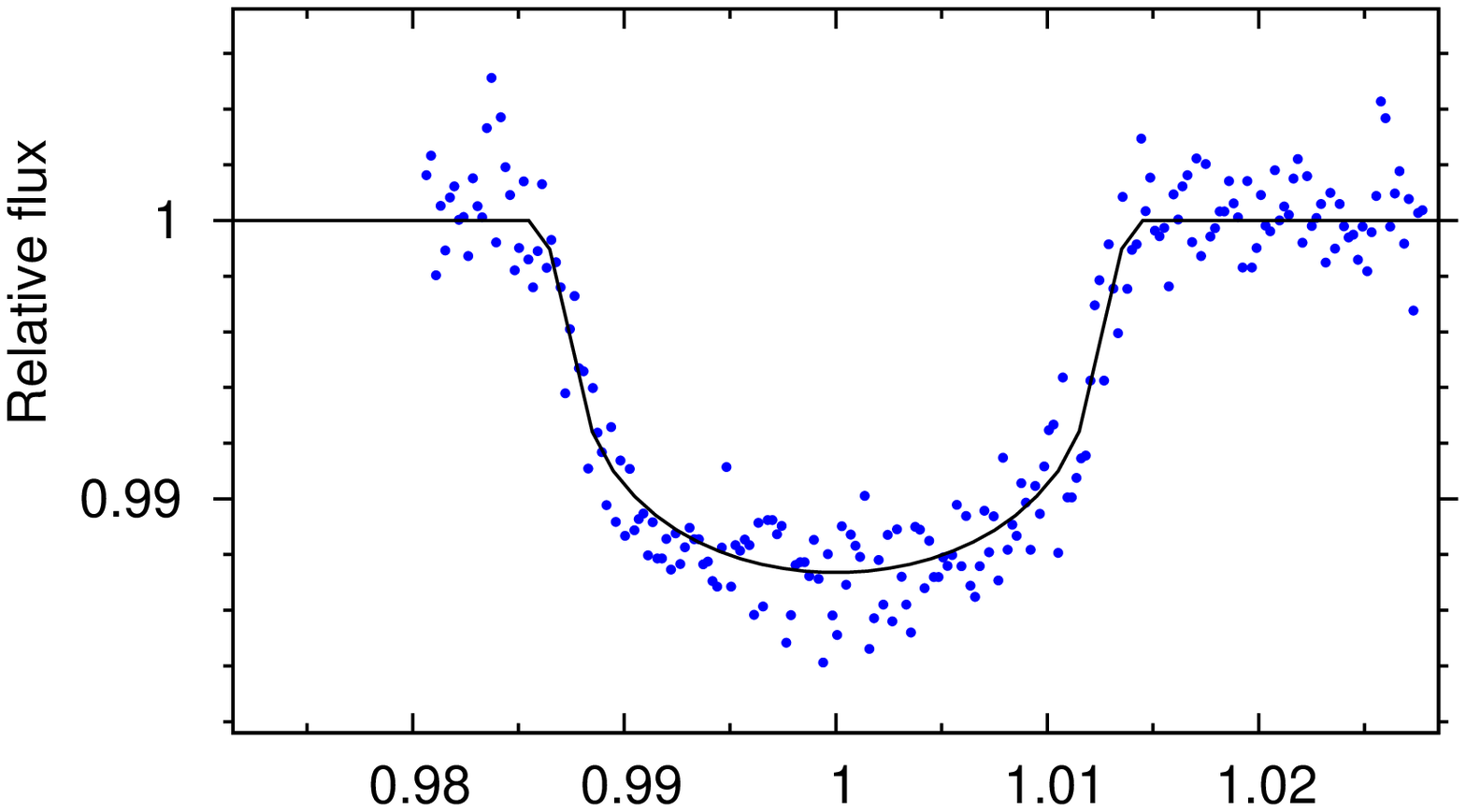}\\ [-2mm]
\hspace*{-5mm}\includegraphics[width=9cm]{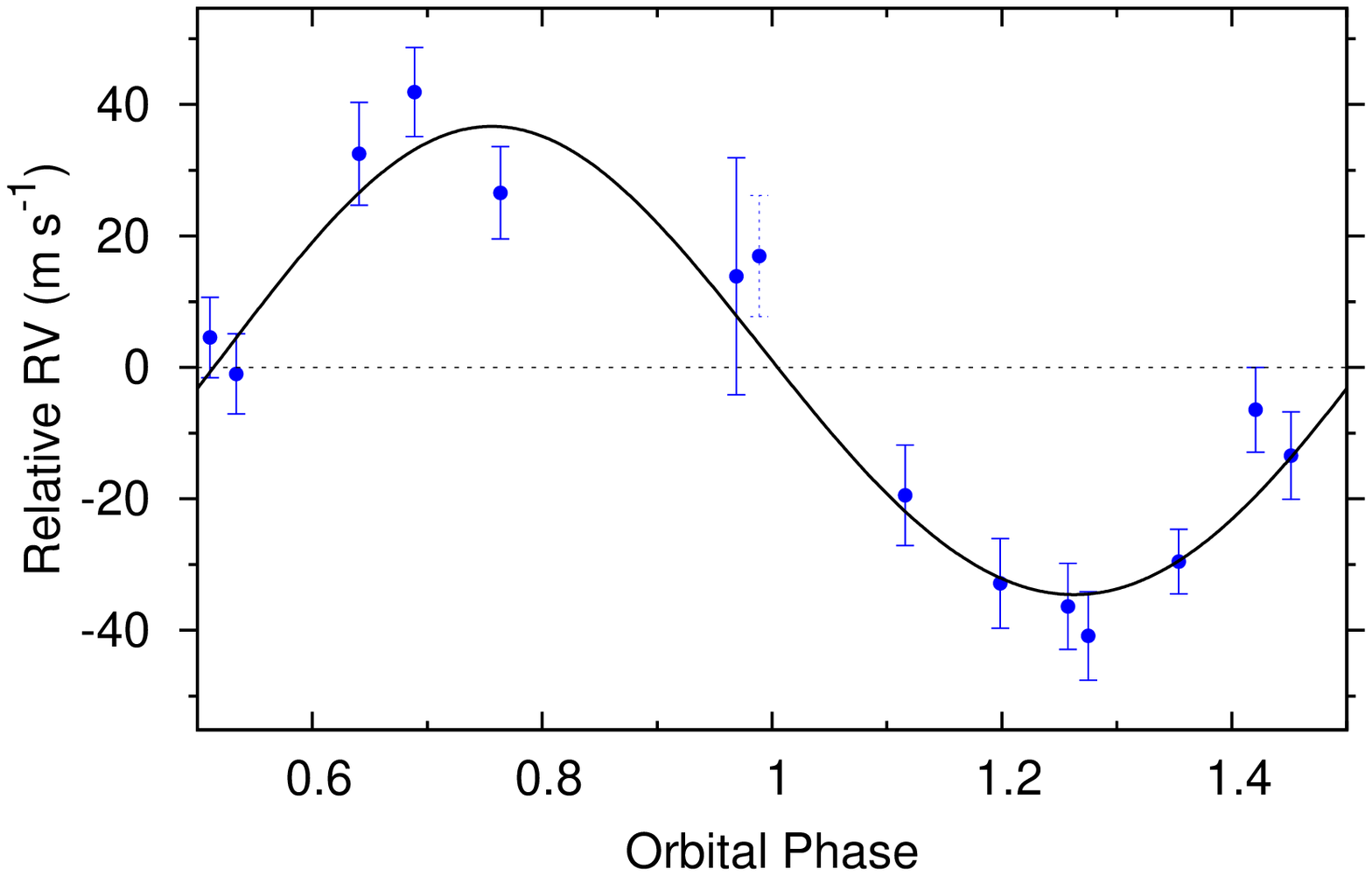}\\ [-2mm]
\hspace*{-5mm}\includegraphics[width=9cm]{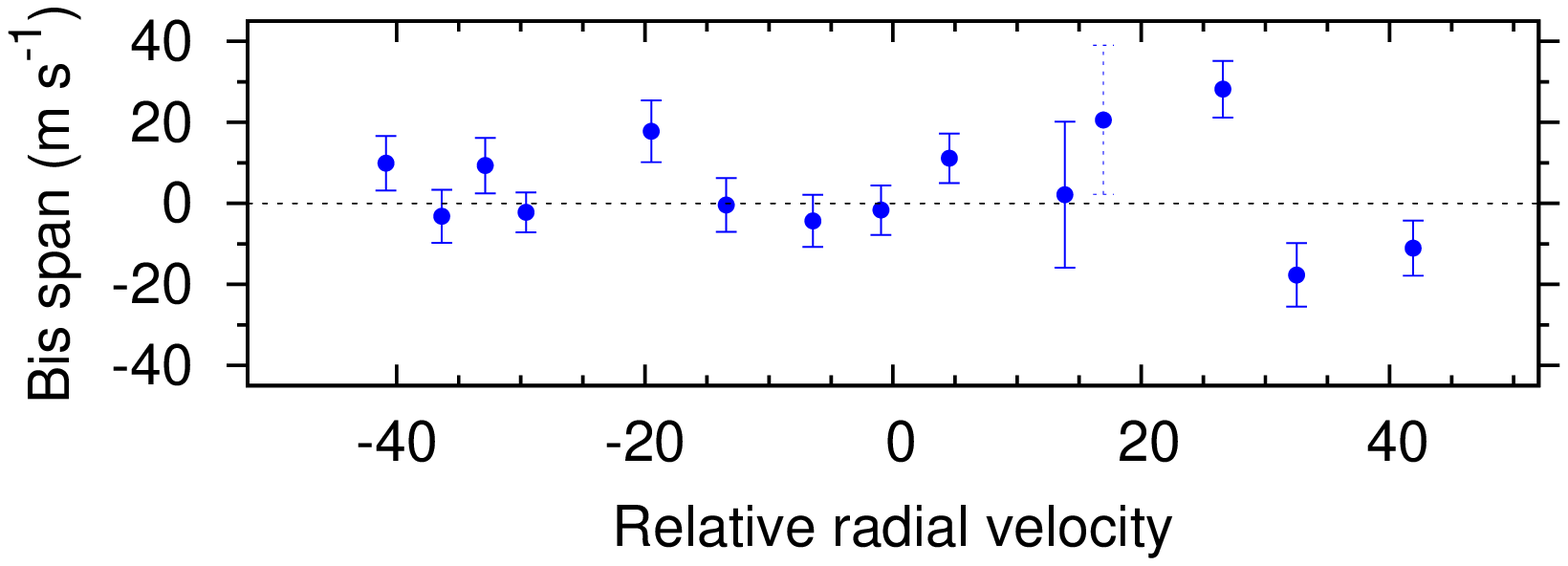}\\ [-2mm]
\caption{(Top) The WASP-South lightcurve folded on the 3.9-d transit
period. (Second panel) The Euler (Gunn r) transit lightcurve with the fitted
MCMC model (phase 1 is HJD 245\,5445.7614). (Third) The CORALIE radial
velocities with the fitted model (the point with dashed error bars was
taken during transit and thus was excluded from the model fit).
(Lowest) The bisector spans; the absence of any correlation with
radial velocity is a check against transit mimics (Queloz et\,al.\
2001).}
\label{figure:wasp}
\end{figure}

\begin{table}
\caption{CORALIE radial velocities of WASP-29.\protect\rule[-1.5mm]{0mm}{2mm}} 
\label{rv-data} 
\begin{tabular*}{0.5\textwidth}{@{\extracolsep{\fill}}cccr} 
\hline 
BJD\,--\,2400\,000 & RV & $\sigma$$_{\rm RV}$ & Bisector\\ 
 & (km s$^{-1}$) & (km s$^{-1}$) & (km s$^{-1}$)\\ [0.5mm]
\hline
55071.8814 & 24.5671 & 0.0068 &\rule{0mm}{5mm} --0.0111  \\
55073.8810 & 24.4924 & 0.0068 & ~0.0093\\
55074.8740 & 24.5118 & 0.0066 & --0.0004\\
55076.9032 & 24.5391 & 0.0180 & ~0.0022\\
55092.6724 & 24.5422 & 0.0092 & ~0.0206\\
55093.7263 & 24.4889 & 0.0066 & --0.0032\\
55094.7205 & 24.5298 & 0.0061 & ~0.0111\\
55095.7121 & 24.5518 & 0.0070 & ~0.0282\\
55097.7176 & 24.4844 & 0.0067 & ~0.0099\\
55098.7330 & 24.5243 & 0.0061 & --0.0017\\
55116.7063 & 24.5058 & 0.0076 & ~0.0178\\
55118.7652 & 24.5577 & 0.0078 & --0.0177\\
55129.6706 & 24.5188 & 0.0065 & --0.0043\\
55168.6361 & 24.4957 & 0.0049 & --0.0022\\
\hline
\multicolumn{4}{l}{Bisector errors are twice RV errors} 
\end{tabular*} 
\end{table}

\section{Observations}
WASP-South is an array of cameras based on 11.1-cm, f/1.8 lenses
which cover a total of 450 square degrees of sky.  The typical 
observing pattern tiles 30-s exposures of several fields with 
a cadence of 8 mins, recording stars in the range $V$ = 8--15. 
The WASP-South survey is described in \cite{sw}
while a discussion of our planet-hunting methods can 
be found in \cite{wasp1}, \cite{wasp3}, and references therein. 

WASP-29 is a $V$ = 11.3, K4V star in the
constellation Phoenix. It was observed by WASP-South 
from May to November in both 2006 and 2007, accumulating 
9161 data points. These data show periodic transits 
with a 3.9-d period (Fig.~1).    There are no other significant
sources within the 48${^{''}}$ extraction aperture 
(3.5 14${^{''}}$ pixels) to dilute the transit depth.  

We used the CORALIE spectrograph on the Euler 1.2-m telescope
at La Silla to obtain fourteen radial-velocity measurements over  
2009 August--December (Table~1).  These show that the 
transiting body is a Saturn-mass planet.   On 2010 September 06
we obtained a transit lightcurve with Euler's CCD camera, using 
20-s, R-band exposures, resulting in a mean
error of 1.5 mmag (Fig.~1).

The CORALIE radial-velocity measurements were combined with the
Euler and WASP-South photometry in a simultaneous Markov-chain Monte-Carlo
(MCMC) analysis to find the parameters of the WASP-29 system 
(Table~2). For details of our methods see  \citet{mcmc} and \citet{wasp3}.  
For limb darkening we used the  4-parameter non-linear law 
of  Claret (2000) with parameters fixed to the values noted 
in Table~2. The eccentricity was a free parameter but the data are compatible 
with a circular orbit.  

One departure from early WASP practice is the way we determine the
stellar mass.  The stellar effective temperature and metallicity are
treated as jump parameters in the Markov chain, and controlled by
Gaussian priors derived from their spectroscopically-determined values
and uncertainties. At each step in the chain the stellar density is
determined from the transit duration and impact parameter. The stellar
mass is then determined at each step as a polynomial function of
$T_{\rm eff}$, [Fe/H] and $\log \rho/\rho_{\odot}$, as determined by 
Enoch et\,al.\ (2010a). This calibration is derived from the
compilation of 40 stars in eclipsing binaries with well-determined
masses, radii, effective temperatures and metallicities, published by
Torres et\,al.\ (2010).

\begin{table} 
\caption{System parameters for WASP-29.\protect\rule[-1.5mm]{0mm}{2mm}}  
\begin{tabular}{lc}
%\begin{tabular}{lc}%{0.5\textwidth}{@{\extracolsep{\fill}}lc} 
\hline
%Parameter (Unit) & Value \rule{0mm}{5mm} \\ [0.5mm] 
%\hline
\multicolumn{2}{l}{Stellar parameters from spectroscopic analysis.\rule[-1.5mm]{0mm}{2mm}} \\
\hline 
%\\ \hline
\multicolumn{2}{l}{RA\,=\,23$^{\rm h}$51$^{\rm m}$31.08$^{\rm s}$, 
Dec\,=\,--39$^{\circ}$54$^{'}$24.2$^{''}$ (J2000)\rule{0mm}{5mm}}\\
\multicolumn{2}{l}{\ (TYC\,8015-1020-1, 2MASS\,J23513108--3954241)}\\
$V$ mag & 11.3  \\ %[2mm]
Spectral type & K4V \\
$T_{\rm eff}$ (K)      & 4800 $\pm$ 150  \\
$\log g$      & 4.5 $\pm$ 0.2 \\
$\xi_{\rm t}$ (km\,s$^{-1}$)    & 0.6 $\pm$ 0.2\\
$v\,\sin i$ (km\,s$^{-1}$)     & 1.5 $\pm$ 0.6 \\
{[Fe/H]}   &  +0.11 $\pm$ 0.14 \\
{[Si/H]}   &  +0.25 $\pm$ 0.08 \\
{[Ca/H]}   &  +0.30 $\pm$ 0.19 \\
{[Ti/H]}   &  +0.38 $\pm$ 0.17 \\
{[Cr/H]}   &  +0.22 $\pm$ 0.16 \\
{[Ni/H]}   &  +0.19 $\pm$ 0.10 \\
log A(Li)  &$<$0.3  \\ [0.8mm] \hline
\multicolumn{2}{l}{Parameters from MCMC analysis.\rule[-1.5mm]{0mm}{5mm}} \\
\hline 
$P$ (d) &  3.922727$\,\pm\,$0.000004 \rule{0mm}{5mm}\\ [0.3mm]
$T_{\rm c}$ (HJD) & 2455320.2341$\,\pm\,$0.004\\ [0.3mm]
$T_{\rm 14}$ (d) & 0.1108$\,\pm\,$0.0015\\ [0.3mm]
$T_{\rm 12}=T_{\rm 34}$ (d) & 0.0108$\,\pm\,$0.0016\\ [0.3mm]
Rel.~Depth (R-band) & 0.0126$\,\pm\,$0.0002\\ [0.3mm] 
$R_{\rm P}^{2}$/R$_{*}^{2}$ &  0.0102$\,\pm\,$0.0004\\ [0.3mm]
$b$ & 0.26$\,\pm\,$0.15 \\ [0.3mm]
$i$ ($^\circ$) & 88.8$\,\pm\,$0.7  \medskip \\ [0.3mm]
$K_{\rm 1}$ (m s$^{-1}$) & 35.6$\,\pm\,$2.7\\ [0.3mm]
$a$ (AU)  & 0.0457$\,\pm\,$0.0006 \\  [0.3mm]
$\gamma$ (km s$^{-1}$) & 24.5252$\,\pm\,$0.0009 \\ [0.3mm]
$e$ &  0.03$^{+0.05}_{-0.03}$ \medskip\\ [0.3mm]
$M_{\rm *}$ (M$_{\rm \odot}$) & 0.825$\,\pm\,$0.033\\ [0.3mm]
$R_{\rm *}$ (R$_{\rm \odot}$) & 0.808$\,\pm\,$0.044\\ [0.3mm]
$\log g_{*}$ (cgs) & 4.54$\,\pm\,$0.04\\ [0.3mm]
$\rho_{\rm *}$ ($\rho_{\rm \odot}$) & 1.56$^{+0.20}_{-0.23}$ \medskip \\ [0.3mm]
$M_{\rm P}$ (M$_{\rm Jup}$) & 0.244$\,\pm\,$0.020\\ [0.3mm]
$R_{\rm P}$ (R$_{\rm Jup}$) & 0.792$^{+ 0.056}_{- 0.035}$ \\ [0.3mm]
$\log g_{\rm P}$ (cgs) & 2.95$\,\pm\,$0.05\\ [0.3mm]
$\rho_{\rm P}$ ($\rho_{\rm J}$) & 0.49$\,\pm\,$0.08\\ [0.3mm]
$\rho_{\rm P}$ (cgs) & 0.65$\,\pm\,$0.10\\ [0.3mm]
$T_{\rm P, A=0}$ (K) & 980$\,\pm\,$40\\ [1mm] 
\hline 
\multicolumn{2}{l}{Errors are 1$\sigma$; Limb-darkening coefficients were:}\\
\multicolumn{2}{l}{a1 = 0.7291, a2 = --0.8130, a3 = 1.5386, a4 = --0.6296}
\end{tabular} 
\end{table} 

\section{WASP-29 stellar parameters}
The 14 CORALIE spectra of WASP-29 were co-added to produce a spectrum
with a typical S/N of 80:1, which we analysed using the methods
described in Gillon et\,al.\ (2009).  We used the H$\alpha$ line to
determine the effective temperature ($T_{\rm eff}$), and the Na\,{\sc
i}\,D and Mg\,{\sc i}\,b lines as diagnostics of the surface gravity
($\log g$). The parameters obtained are listed in Table~2. The
elemental abundances were determined from equivalent-width
measurements of several clean and unblended lines. A value for
microturbulence ($\xi_{\rm t}$) was determined from Fe\,{\sc i} using
Magain's (1984) method. The quoted error estimates include that given
by the uncertainties in $T_{\rm eff}$, $\log g$ and $\xi_{\rm t}$, as
well as the scatter due to measurement and atomic data uncertainties.

The temperature and $\log g$ values are consistent with a K4 main-sequence
star, and this is also consistent with the {\sl BVRIJHK\/} magnitudes collected by SIMBAD.  There is some indication of above-Solar metal abundances (Table~2). 

The projected stellar rotation velocity ($v\,\sin i$) was determined
by fitting the profiles of several unblended Fe\,{\sc i} lines. We
assumed a value for macroturbulence ($v_{\rm mac}$) of 0.5 $\pm$ 0.3
km\,s$^{-1}$, based on the tabulation by Gray (2008), and an
instrumental FWHM of 0.11 $\pm$ 0.01 \AA, determined from the telluric
lines around 6300\AA. The best-fitting value of $v\,\sin i$ was 1.5
$\pm$ 0.6 km\,s$^{-1}$.

\subsection{Evolutionary status of WASP-29}
The temperature and density of the host star WASP-29 are shown
on a modified H--R diagram in Fig.~2. The best-fit values place
it above the ZAMS, which would indicate either a pre- or 
post-main-sequence star, while the absence of lithium and 
the low value of $v\,\sin i$ of 1.5\,$\pm$\,0.6 km\,s$^{-1}$ 
argue for the latter.  Plotting against evolutionary tracks from 
Demarque et\,al.\ (2004), and using a metallicity of [M/H] = 
0.2, near the mean of the values in Table~2, indicates an 
age of 15 Gyr with a 1-$\sigma$ lower limit of 7 Gyr (and an
upper limit beyond the 20-Gyr oldest isochrone).  
For this metallicity, track fitting results in a stellar 
mass of 0.78$\,\pm\,$0.05 M$_{\odot}$, compatible with 
the 0.83\,$\pm$\,0.03 M$_{\odot}$ derived in Section~2.  

Increasing the metallicity to [M/H] = 0.3 reduces the age to 
12 Gyr, with a 1-$\sigma$ lower limit of 5 Gyr, while reducing 
the metallicity would increase the age, with [M/H] = 0 giving 
an age of 20 Gyr. 

For a K4V star, $V$ = 11.3 would indicate a distance of $\sim$\,70 pc.
The proper motion of 0.1$^{\prime\prime}$\,yr$^{-1}$ (Zacharias et\,al.\ 2004) 
then indicates a transverse velocity of 33 km\,$^{-1}$, which,
with our measured radial velocity of 24.5 km\,$^{-1}$, gives 
a space velocity of 40 km\,$^{-1}$ relative to us, which is typical of
a local thin-disk star (e.g.\ Navarro et\,al.\ 2010). 

Thus the properties of WASP-29 are compatible with a local
thin-disk star, provided that its metallicity is above Solar
and that its age is towards the younger
end of the current error range, thus bringing it within the
$\sim$\,9 Gyr age of the thin disk.   For
this reason it will be worthwhile to obtain better
parametrizations of WASP-29's metallicity and effective temperature.

 \begin{figure}
\hspace*{-5mm}\includegraphics[width=9cm,angle=-90]{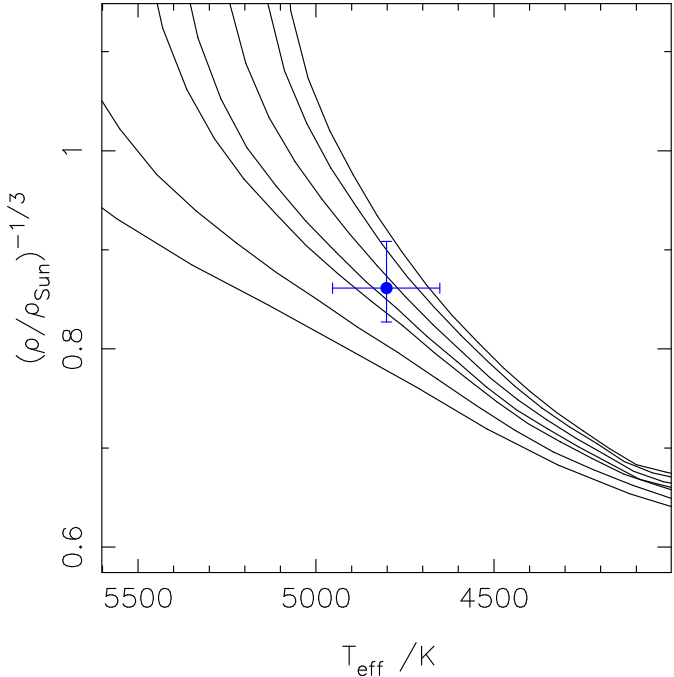}
\caption{Evolutionary tracks on a modified H--R diagram
($\rho_{*}^{-1/3}$ versus $T_{\rm eff}$).  The isochrones are (in
order from the left) 1, 5, 10, 12, 15, 18 \&\ 20 Gyr, for a
metallicity of [M/H]\,=\,+0.25 (from Demarque et\,al.\ 2004).}
\end{figure}

\section{Discussion}
We show in Fig.~3 the mass--radius distribution for 
transiting exoplanets.  WASP-29b is now the planet closest
in mass and radius to Saturn itself, and among the Saturn-mass
planets is midway between the dense CoRoT-8b (Bord\'e et\,al.\ 2010)
and the more bloated systems WASP-21b (Bouchy et\,al.\ 2010)
and HAT-P-12b, HAT-P-18b and HAT-P-19b (Hartman et\,al.\ 2009, 2010). 

The smaller radius of WASP-29b among Saturn-mass exoplanets
is unlikely to be caused primarily by irradiation, since the three 
HAT planets and WASP-29b all orbit stars of K1V to K4V, while 
WASP-29b has the shortest orbital period and so will be the
most irradiated.  Further, the irradiation for HD\,149026b
is greater, yet it is denser. In addition, all of the above
planets have eccentricities compatible with zero,
suggesting that tidal heating is not currently important. 

It has been suggested that metallicity is a major 
factor in determining the radii of Saturn-mass planets
(Hartman et\,al.\ 2009; Bouchy et\,al.\ 2010), with
higher-metallicity systems having larger cores and thus
smaller radii. WASP-29 is in
line with this pattern, with indications of an 
elevated abundance of iron, [Fe/H] = +0.11\,$\pm$\,0.14, 
and other metals (Table~2).  From the theoretical 
models of Fortney et\,al.\ (2007) and
Baraffe et\,al.\ (2008), WASP-29b could have  
a heavy-element core  of approximately 25 M$_{\oplus}$ compared 
to $\sim$\,50 M$_{\oplus}$
for the denser HD\,149026b (Carter et\,al.\ 2009), and contrasting
with $<$\,10 M$_{\oplus}$ for the less-dense HAT-P-12b 
(Hartman et\,al.\ 2009) and WASP-21b (Bouchy et\,al.\ 2010).    
However, the recently 
announced planets HAT-P-18b and HAT-P-19b 
break the pattern by being under-dense while having
above-solar metallicities (Hartman et\,al.\ 2010). 
Thus, although there does appear to be an overall
correlation between metallicity, irradiation and 
planet radii (Enoch et\,al.\ 2010b; Anderson et\,al.\ 2010)
these factors cannot be the full explanation.

It is worth remarking that the known transiting Saturns 
mostly orbit K dwarves, with all the stars except HD\,149026 being G7 or later.
This is likely a selection effect, since transits of smaller 
planets are easier to detect against smaller stars
(the exception, HD\,149026b, was first found by radial-velocities; 
Sato et\,al.\ 2005).    
Similarly, while radial-velocity surveys find more
planets around higher-metallicity
stars (Santos et\,al.\ 2004), a correlation of lower 
metallicity with larger planet radius would
bias transit-survey detections to lower-metallicity systems.

\begin{figure}
\hspace*{-5mm}\includegraphics[width=9cm]{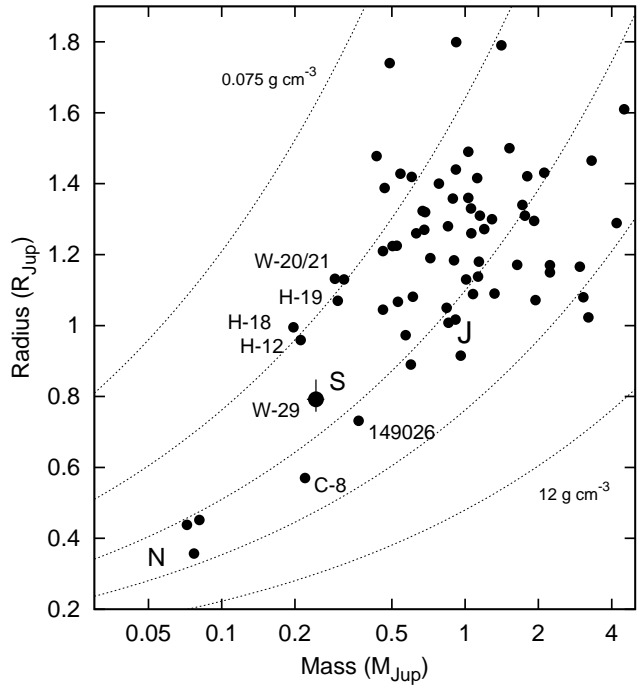}
\caption{The mass--radius plot for transiting exoplanets. 
W = WASP, H = HAT, C = CoRoT, 149026 = HD\,149026b, and 
the symbols J, S \& N mark the location of Jupiter, Saturn 
and Neptune. Dotted lines are density contours (at 12, 3, 1, 0.3 \&\ 0.075 g\,cm$^{-3}$). Data from Schneider (2010, as of August).}
\end{figure}

\acknowledgments
%\section*{Acknowledgments}
WASP-South is hosted by the South African
Astronomical Observatory and we are grateful for their ongoing
support and assistance. Funding for WASP comes from consortium universities
and from the UK's Science and Technology Facilities Council.

\end{document}